\documentclass[prl,twocolumn]{revtex4}
\usepackage{color}
\usepackage{graphicx}
\usepackage{subfigure}
\usepackage{amsmath}
\usepackage{amsfonts}

\begin{document}

\title{Non-Equilibrium Response of Deuterium-Tritium Fusion Plasma: Molecular Dynamics Study of Ion Tail Replenishment}
\date{September 9, 2013}

\author{Jonathan L. Belof}
\email{belof1@llnl.gov}
\affiliation{Lawrence Livermore National Laboratory\\7000 East Avenue., Livermore, CA 94550\\}

\author{Scott Wilks}
\affiliation{Lawrence Livermore National Laboratory\\7000 East Avenue., Livermore, CA 94550\\}

\author{Jose O. Sinibaldi}
\affiliation{Lawrence Livermore National Laboratory\\7000 East Avenue., Livermore, CA 94550\\}

\begin{abstract}

It has recently been proposed that transport of energetic deuterium-trition ions (Knudsen layer), owing to their long mean free path, may deplete the tail of the velocity distribution in the vicinity of the Gamow peak and thus serve to greatly reduce the yield of an inertial confinement fusion (ICF) plasma [Molvig \emph{et al.}, \emph{Phys. Rev. Lett.}, 109:095001 (2012)].  Molecular dynamics simulations of the non-equilibrium response of a fusion plasma to ion distribution tail loss show that equilibrium is re-established extremely rapidly, with ion tail replenishment complete in less than 1 ps for $\rho r > 0.01$.  The exceedingly fast relaxation of the fusion plasma dynamics, relative to the Fokker-Planck collisional operator, can be understood in terms of a large entropic driving force that pushes the phase space distribution toward a Maxwellian, and is a characteristic of the Lyupanov instability for the Coulomb system.  The thermal reactivity $\langle \sigma v \rangle$ for the D(t,n)$\alpha$ fusion reaction is calculated as a function of progress toward equilibrium for a series of densities relevant to ICF hot spot ignition.

\end{abstract}

\maketitle

The creation of a self-sustained thermonuclear burn has been the goal of research in inertial confinement fusion (ICF) for decades.  One approach is to form a high density ``hot spot'' within a deuterium-tritium (DT) plasma approximately 50 $\mu$m in size, with $\rho \approx$ 300 g/cc and $T \approx$ 10 keV.  Hot spot ions in the high energy tail of the velocity distribution, around the Gamow peak, may tunnel through the Coulomb barrier and fuse.  The stopping of 3.5 MeV $\alpha$-particles then rapidly heats the electron field which will eventually heat the DT ions, through electron-ion coupling, to a very high temperature of 50-100 keV and, in principle, propagate a burn wave.\cite{lindl}

The high energy density conditions described above are formed under hydrodynamic (inertial) compression, and held for a confinement time $\tau_{conf}$, commonly given by the Lawson criteria\cite{atzeni}, in order to initiate burn prior to capsule disassembly.  As the fusion process proceeds, the high energy ion tail of the burning DT would become depleted, were it not for the extremely rapid replenishment of the tail -- an insight that has been discovered through Fokker-Planck particle simulations.\cite{sherlock,michta}  The topic of this letter deals with the question of whether or not replenishment of the high energy tail of the hot spot ions may occur sufficiently fast (\emph{i.e.} $\tau_{eq} \ll \tau_{conf}$) so as to negate a Knudsen layer effect.

In past years, it has been proposed that the high energy tail of the hot spot ions might become depleted through non-local transport\cite{henderson,petschek,nishikawa}, and this mechanism has been expanded upon in the recent letter by Molvig \emph{et al.}\cite{molvig} They propose that the velocity distribution of the hot spot ions will become non-Maxwellian since the high energy ions have a much longer mean free path than the thermal ions, owing to the Coulomb cross section scaling as $\sigma_{coul} \propto \frac{Z^2}{E^2}$, and that these ions will be lost due to subsequent collisions with a non-reactive ``wall'' (ablator, cold DT fuel, \emph{etc.}).  This effect is expressed non-locally in terms of a ``Knudsen number'' (in analogy with molecular kinetic theory), $N_k$, that characterizes the distance from the wall relative to the mean free path of the high energy tail, with ions travelling a distance $L$ having achieved system escape.  Ion tail depletion (represented by a diffusion operator) is taken as a loss term, with collisional energy upscatter (approximated as Coulomb scattering through a static Maxwellian field) presenting a source term -- the steady state solution is the Knudsen distribution function.  Consequently, the energetic tail in the vicinity of the Gamow peak would be less occupied, reducing the ICF yield substantially.

However, there is an additional source term that has not been considered in the analytic approximations used to derive the Knudsen layer result: replenishment of the tail due to the non-equilibrium response of the depleted system.  Indeed, the seminal work on the subject by Henderson\cite{henderson} states: ``In cases in which these conditions are met, we need to consider the rate at which collisions are able to fill out the distribution function. This requires a Fokker-Planck calculation which is not in hand.''  It is clear that even in the absence of consideration for the upscatter source term, there will be a restoring force acting on the velocity distribution in the thermodynamic direction toward a Maxwellian.

The Knudsen distribution function is derived in Ref. \cite{molvig} as
\begin{eqnarray}
f_k(v) = \frac{2}{\sqrt{\pi + N_k \left( \beta \frac{1}{2} m v^2 \right)^{\frac{3}{2}} }} e^{-\beta \frac{1}{2} m v^2 - \frac{2}{5} N_k \left( \beta \frac{1}{2} m v^2 \right)^{\frac{5}{2} } } \label{eq:knudsen_distribution}
\end{eqnarray}
where $\beta = 1/k_b T$ is the inverse temperature.  Eq. \ref{eq:knudsen_distribution} results from application of the WKB approximation, and assumes a particular form for the diffusion operator.  This approximate form for the Knudsen velocity distributon function given by Eq. \ref{eq:knudsen_distribution} assumes the least conservative case for the diffusion operator, which was justified in Ref. \cite{molvig} as simply providing an estimate of the effect.  For example, for a DT plasma with $\rho$ = 6 g/cc and $T$ = 9 keV (a state point examined in both Ref. \cite{molvig} and here) and $N_k$ = 0.1 the reduction in fusion yield is a factor of $\approx 4 \times$; improvements in the derivation of the Knudsen distribution (\emph{e.g.} approximations less severe than WKB, \emph{etc.}) all serve to lessen the yield reduction effect.\cite{zimmerman}  Nonetheless, we perform all of our comparisons here with Eq. \ref{eq:knudsen_distribution} so that our assessment in terms of relaxation time will be the most conservative possible.

Here we have turned to first principles simulation -- molecular dynamics with Coulomb scattering -- to address the question of whether neglecting tail replenishment can be justified.  It is the significant finding of this letter that in fact the tail of the hot spot DT ions is replenished on the sub-picosecond timescale, $>1000 \times$ faster than the confinement time $\tau_{conf} \approx$ 1 ns, thus reducing the possible width of a Knudsen layer substantially.  For example, if we assume the Gamow peak (30 keV) ions to be travelling toward escape ballistically, then in 1 ps they will have travelled only 1.5 $\mu$m by which time the tail has already been replenished (with a slightly cooler thermal ion temperature remaining).  A similar estimate is obtained using the mean free path from spitzer\cite{spitzer}, $\lambda = m^2 v^4 / 8 \pi n e^4 \ln \Lambda$. \emph{E.g.}, for $\rho=6$ g/cc, $T=9$ keV, $N_k=0.1$ and $L=10 \mu$m there would be almost no yield reduction, in stark contrast with previously quoted\cite{molvig} $4\times$ reduction values.  Furthermore, for densities more typical of an ideal fusion hot spot, $\rho=100$ g/cc and $T=9$ keV, the equilibration time is found to be an extremely fast 100 fs (!) as a result of the increased collisional probability that scales with density.

As a first estimate, the plasma equilibration time is given by\cite{zeldovich}
\begin{eqnarray}
\frac{1}{\tau_{eq}} \approx \frac{2\pi}{9} N \overline{v} \beta^2 Z^4 \ln \Lambda \label{eq:tau_zeldovich}
\end{eqnarray}
with $\overline{v} = \sqrt{8/\pi m \beta}$ the mean thermal velocity, and where the equilibration time $\tau_{eq}$ is defined by Zel'dovich as the time required for the system to undergo enough collisional processes such that the average change in momentum is comparable to the momentum value itself.  For the illustrative case of DT plasma with $\rho$ = 6 g/cc, $T$ = 9 keV and $\ln \Lambda$ = 6, we estimate a relaxation time $\tau_{eq} \approx$ 1.5 ps.  We note that Eq. \ref{eq:tau_zeldovich} is derived for a 2-body, short ranged Coulomb impact -- effects from many-body collisions (which will occur in a true plasma due to density fluctuations) and long-range Coulomb interactions will only serve to reduce $\tau_{eq}$ from that given by Eq. \ref{eq:tau_zeldovich}, and it is precisely these effects that we determine through molecular dynamics (MD).  Furthermore, an assumption of $\ln \Lambda$ is not required since the ion collisions are resolved explicitly \emph{via} the particle dynamics.

We consider a system of $N$ particles that are initially out-of-equilibrium at time $t=0$.  The initial state is taken to be the Knudsen distribution $f_k$, Eq. \ref{eq:knudsen_distribution}, which will eventually relax to a Maxwellian distribution of temperature $T_{eq}$ at time $t=\tau_{eq}$.  By equipartition and conservation of energy, the final equilibrium temperature will be
\begin{eqnarray}
T_{eq} = \frac{2}{3 N k_b} E_{neq} \label{eq:T_eq}
\end{eqnarray}
where $E_{neq}$ is the energy of the (initially) non-equilibrium system of $N$ particles, and $T_{eq}$ is the temperature of the Maxwellian that is established after relaxation.  In principle, $E_{neq}$ could be evaluated at the first timestep of the molecular dynamics simulation only and then presumed to be conserved throughout the trajectory -- in reality the numerics lead to small amounts of energy drift in the MD trajectory, and so instead we extract $E_{neq}$ at each timestep (as explained further below).

\begin{figure}[t]
\includegraphics[width=3.5in]{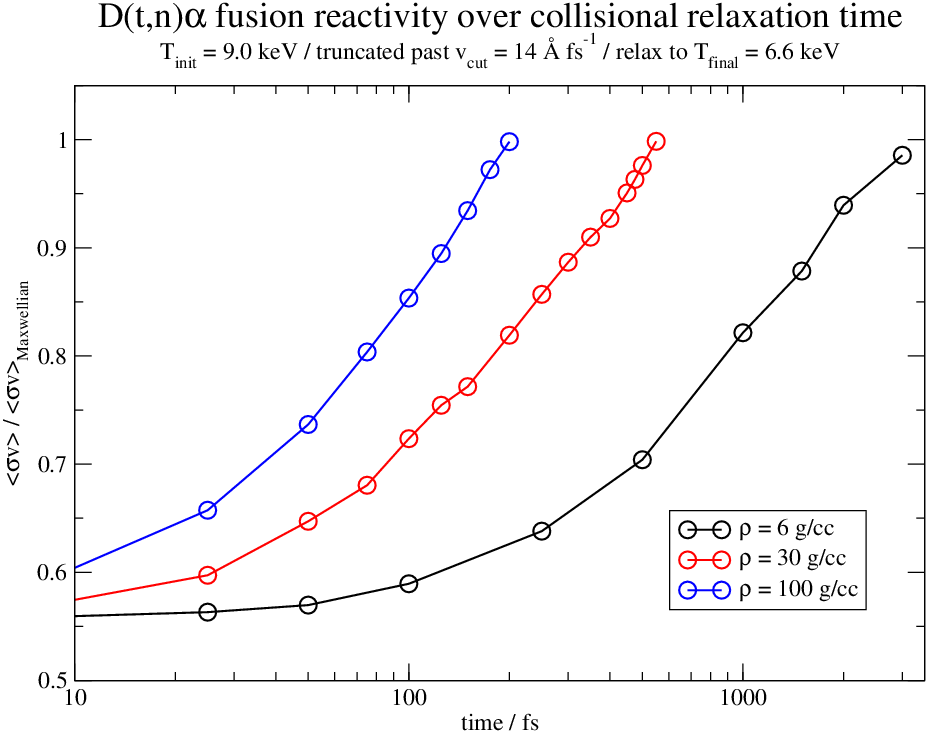}
\caption{Thermal fusion reactivities of deuterium-tritium computed from molecular dynamics simulation of ion distribution function relaxation.  The velocity distribution functions were truncated past the Gamow peak, at 14 $\textrm{\AA}$/ps and then allowed to relax to equilibrium in a time-dependent manner.  As a function of time, the ion distribution function was integrated with the Bosch-Hale cross sections to obtain the thermal fusion reactivity.  The ratio of the time-dependent thermal fusion cross-section to that of a Maxwellian is plotted and represents a metric for which suppression of fusion rates due to Knudsen layer effects is reduced.}
\label{fig:sv}
\end{figure}

Our metric for evaluating the approach to equilibrium
\begin{eqnarray}
\gamma(t) = \frac{\int\limits_0^\infty dv \, f(v,t) f(v; T_{eq})}{\int\limits_0^\infty dv \, |f(v,t)|^2} \label{eq:gamma}
\end{eqnarray}
represents the overlap between the arbitrary non-equilibrium velocity distribution, $f(v,t)$, and the Maxwellian distribution, $f(v;T_{eq})$ corresponding to the system at the equilibrated temperature $T_{eq}$.  We note that at $t=0$ we have $f(v,t=0)=f_k$, and at equilibrium ($t=\tau_{eq}$) we have $\gamma(\tau_{eq}) = 1$, signifying the establishment of a Maxwellian velocity distribution.  We note that the standard sources of MD error, energy drift due to finite order integration and long range force accuracy, will have the effect of heating the simulated system very slightly, by less than 1 \%.  We account for this fact in the calculation of $\gamma(t)$ by using the current energy at each timestep in the determination of $T_{eq}$ in Eq. \ref{eq:T_eq} rather than assuming it to be constant throughout the MD trajectory.

Constant-energy molecular dynamics simulations utilized the velocity Verlet time integrator, with a timestep of $dt=0.0001$ fs, and exhibited good energy conservation with energy drift of less than 85 eV out of 9 keV within 3.0 picoseconds.  A system size of $N=64,000$ particles ensured both adequate statistics of the tail distribution as well as an accurate treatment of the periodic boundary conditions.  Each particle was assigned a mass of 2.5 g/mol, the average mass of the deuterium-tritium mixture.  The Coulomb interactions were calculated with the Particle-Particle-Particle-Mesh (PPPM) Ewald method to an accuracy better than $10^{-8}$ eV/$\textrm{\AA}$ in the forces.  A mean field approximation of the electrons has been applied, whereby the $N$ electrons are represented by a uniform charge neutralizing background.  All calculations were performed with the MD simulation code LAMMPS\cite{lammps}

The inital stage of MD simulation began with random particle positions and velocities assigned by random sampling of a Maxwellian, and then run for several picoseconds to ensure an equilibrated system.  After stopping the simulation, the velocity vectors were reassigned from a random sampling of the Knudsen distribution, Eq. \ref{eq:knudsen_distribution}.  The simulation was then restarted and the trajectory calculated for 3.0 ps, with the equilibration function $\gamma(t)$ evaluated over this trajectory.  The relaxation time $\tau_{eq}$ was then determined based upon the asymptotic approach of $\gamma(t) \rightarrow 1$.

\begin{figure}[t]
\includegraphics[width=3.5in]{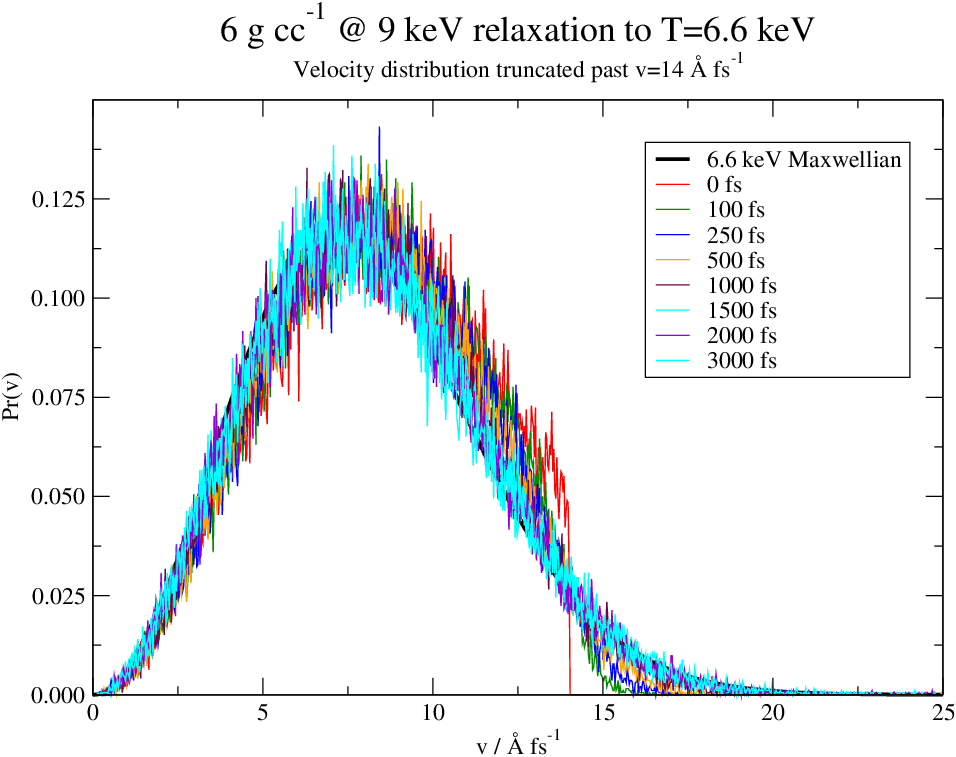}
\caption{Evolution of the ion velocity distribution after truncation in the vicinity of the Gamow peak.}
\label{fig:vdist}
\end{figure}

Scaling studies were undertaken whereby the approach to equilibrium, determined by Eq. \ref{eq:gamma}, was calculated with respect to variation in (a) particle system size $N$, (b) integrator timestep $dt$, and (c) PPPM accuracy $\delta f_{pppm}$.  The asymptotic value of $\tau_{eq}$ was then determined by extrapolation toward the exact limits for each case ($N \rightarrow \infty$, $dt \rightarrow 0$, $\delta f_{pppm} \rightarrow 0$).  The results are shown in Fig. \ref{fig:vdist} for DT $\rho$ = 6 g/cc and $T$ = 9 keV, with a relaxation time of 250 femtoseconds.  We further note that this entire study is more conservative than necessary, since the initial state being a Knudsen distribution is unphysical -- the tail would be replenished continually as the system is compressed and heated, and would merely result in a slightly lower thermal temperature.  A final remark is that we've taken the implicit assumption of complete system escape of the high energy tail (\emph{i.e.}, a perfectly absorbing wall) in the Knudsen layer theory at face value, and have not considered the possibility of interactions with the wall partially reflecting ions back into the reacting system.

The origin of the extremely rapid dynamics leading to tail replenishment can be elucidated through the entropy production.  If we take the non-Maxwellian system to be in a steady state, then it is meaningful to define a local entropy.  The entropy production from $t=0$ to $t=\tau_{eq}$ can then be written using the Gibbs entropy
\begin{eqnarray}
\Delta S_{eq} = -k \left[ f(v;T_{eq}) \ln f(v;T_{eq}) -  f_k \ln f_k \right] \label{eq:gibbs_entropy}
\end{eqnarray}
$T_{eq}$ is calculated \emph{via} Eq. \ref{eq:T_eq} with the system energy $E_{neq}$ determined by integrating the Knudsen distribution $E_{neq} = \int dv \, \frac{1}{2} m v^2 f_k(v)$.   Since there is no free energy barrier to equilibration for this special case of velocity relaxation, we may make the \emph{ansatz} that the equilibration rate is
\begin{eqnarray}
\dot{\gamma(t)} \propto \frac{1}{\tau_{eq}} \exp\left\{ - \frac{\Delta S_{eq} t}{k_b} \right\} \label{eq:gammadot}
\end{eqnarray}
The entropic force driving the plasma toward thermodynamic equilibrium is significant and illustrates the importance of it's consideration along side the purely energetic aspects of a Coulomb system driven out of equilibrium.  An interesting future study would be an analysis of the Lyupanov instability\cite{evans} of high energy density plasmas.

Lawrence Livermore National Laboratory is operated by Lawrence Livermore National Security, LLC, for the U.S. Department of Energy, National Nuclear Security Administration under Contract DE-AC52-07NA27344.

\end{document}